\documentclass[10pt,letterpaper]{article}
\usepackage{opex3}

\begin{document}

%%%%%%%%%%%%%%%%%% title page information %%%%%%%%%%%%%%%%%%
\title{Breaking a quantum key distribution system through a timing side channel}

\author{Ant\'{\i}a Lamas-Linares and Christian Kurtsiefer}

\address{Department of Physics, National University of
Singapore\\2 Science Drive 3, Singapore 117542}
\email{antia\_lamas@nus.edu.sg} %% email address is required
\homepage{http://qoptics.quantumlah.org/lah/} %% author's URL, if desired


\begin{abstract}
The security of quantum key distribution relies on the validity of
quantum mechanics as a description of nature and on the
non-existence of leaky degrees of freedom in the practical
implementations. We experimentally demonstrate how, in some
implementations, timing information revealed during public
discussion between the communicating parties can be used by an
eavesdropper to undetectably access a significant portion of the
``secret'' key.
\end{abstract}

\ocis{(030.5260) Coherence and statistical optics: photon counting; (270.5290) Quantum optics: photon statistics; (999.9999) Quantum cryptography} % REPLACE WITH CORRECT OCIS CODES FOR YOUR ARTICLE

%%%%%%%%%%%%%%%%%%%%%%% References %%%%%%%%%%%%%%%%%%%%%%%%%
%\bibliography{universal}
%\bibliographystyle{osajnl}

%%%%%%%%%%%%%%%%%%%%%%%%%%  body  %%%%%%%%%%%%%%%%%%%%%%%%%%
\section{Introduction}
Theoretical proofs of the security of quantum key distribution
(QKD), are a well developed subfield in quantum communication
research (see~\cite{dusek:06}), both in highly
idealized~\cite{bennett:84, ekert:91} and more realistic
scenarios~\cite{gisin:07}. By construction, these proofs assume that
the legitimate parties measurement results are isolated from the
environment and thus from an eavesdropper. Comparatively little work
has been done studying the possible physical side channels
associated with particularities of the physical devices
used~\cite{kurtsiefer:02, kurtsiefer:01a} or possible attacks based
on the external manipulation of the expected response of the
apparatus~\cite{makarov:05, makarov:06, gisin:06, zhao:07}.

%\section{Timing side channel}
All photon-counting implementations of QKD identify a signal photon
from background by measurement of the arrival time at detectors.
 In an ideal scenario, there can be no correlation between the measurement
outcome on the quantum variable (e.g. polarization in the original
BB84 proposal), and this publicly exchanged timing information.
However, in a recent entanglement based QKD implementation, a pulsed
down-conversion source provided photon pairs with a well-defined
timing signature~\cite{ursin:06}. For photon identification, timing
information was recorded with a high resolution and communicated to
the other side (similar scheme as
in~\cite{kurtsiefer:02,poppe:04,resch:05,marcikic:06}). We show that
there may be an exploitable correlation between the exchanged timing
information and the measurement results in the quantum channel.

\section{Time response analysis}
\begin{figure}
\begin{center}
\includegraphics[scale=0.85]{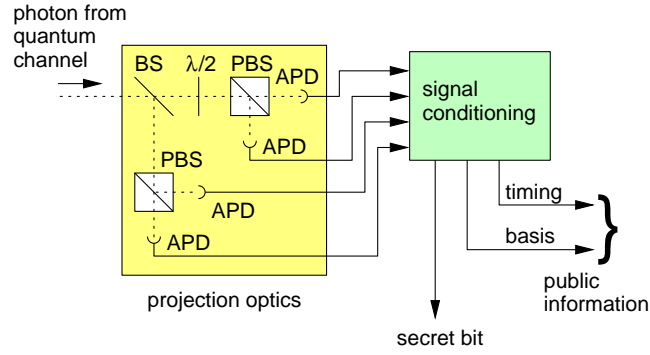}
\end{center}
\caption{A typical configuration of photocounting detectors for
quantum key distribution. A beam splitter (BS), polarizing beam
splitters (PBS) and a half wave plate ($\lambda/2$), divert incoming
photons onto a set of detectors, which generate a macroscopic timing
signal. This timing information and e.g. a projection basis is
revealed publicly, while information on which detector out of two
absorbed a photon is the secret used to subsequently generate a
key.} \label{fig:topology}
\end{figure}

\begin{figure}
\begin{center}
\includegraphics{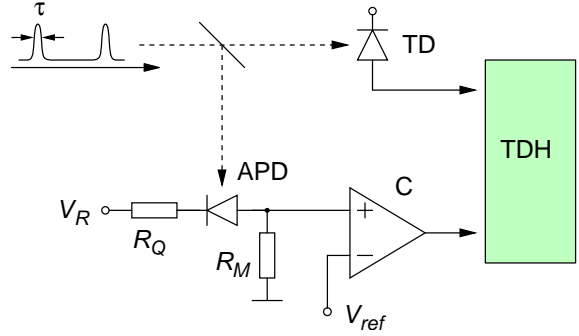}
\end{center}
\caption{Experimental set-up to characterize the timing jitter of a
single
  photon detector. A train of ultrashort light pulses from a mode-locked
  Ti:Sapphire laser is sent with strong attenuation into
  a passively quenched Si avalanche photodiode (APD). A histogram of timing differences (TDH) with respect to the signal of a
  trigger photodidode (TD) is recorded.
} \label{fig:expsetup}
\end{figure}

A configuration implementing the detection scheme just described is
shown in Fig.~\ref{fig:topology}. An incoming photon is randomly
directed by a beam splitter towards two possible polarizing beam
splitters each of which performs a measurement in one basis ($H/V$
or $45^\circ/-45^\circ$). Finally, there are four possible outcomes
of the measurement (two bits of information) of which one bit will
be made public. The remaining bit is the raw material for generating
the secret key and must be kept secret. Although the optical
distance from the entrance of the module to the four detectors
differs by less than 1\,mm, there is a measurable difference in the
timing of the electronic signal from the different possibilities.
In order to determine the timing differences between the four single
photon detectors, we used an attenuated fraction of a pulse train
emitted by a Ti:Sapphire femtosecond laser as a light source (see
Fig.~\ref{fig:expsetup}). Single photon detectors consisted of
Silicon Avalanche Photodiodes (type C30902S, Perkin-Elmer), operated
in a passively quenched configuration. The breakdown of the
avalanche region was converted into a digital pulse signal by a high
speed comparator, registering a voltage drop over the measurement
resistor $R_M=100\,\Omega$ of 150\,mV, which has to be compared to a
maximal voltage drop across $R_M$ of about 700\,mV. The distribution
of peak amplitudes for the breakdown signal exhibits a spread below
10\% for photodetector event rates of 5000--6000\,s$^{-1}$, and the
pulse duration before the comparator is on the order of 2\,ns.

\begin{figure}
\begin{center}
\includegraphics{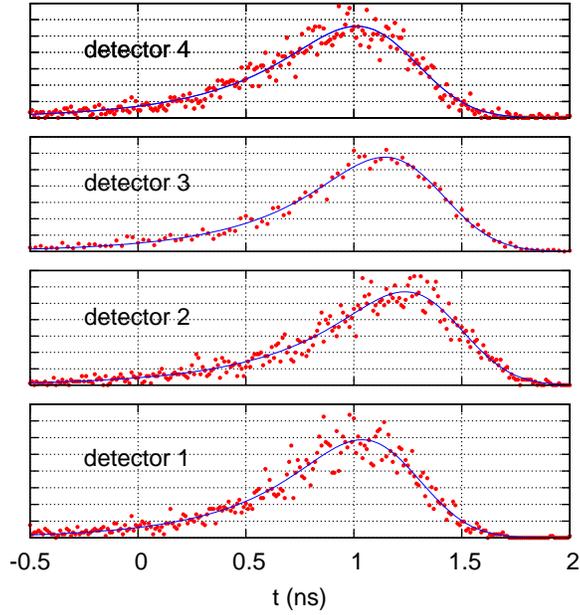}
\end{center}
\caption{Photoevent timing histograms for the four detectors
involved in a
  quantum key distribution receiver. While the general shape of the
  distributions is similar, there is a distinction in the response time
  visible for detectors 1 and 4 with respect to detectors 2 and 3, which, if
  not compensated,  can be exploited by an eavesdropper to gain knowledge
  about the measurement result. The solid lines represent a fit to the model
  in equation \ref{eq:modelfunction}.
} \label{fig:histograms}
\end{figure}

We obtained the timing distribution with an oscilloscope sampling at
20\,GS/s, by interpolating the time when the comparator output
passed through the 50\% value between the two logical levels. Time
reference is a trigger signal supplied by a MSM Schottky reference
photodiode (G7096-03, Hamamatsu) looking at another fraction of the
optical pulse train.  The timing jitter of 10\,ps (FWHM) we observe
between consecutive pulses from the mode-locked laser gives an upper
bound for the total timing uncertainty.
The resulting timing histograms of the different detectors
(Fig.~\ref{fig:histograms}) show a clearly different centroid
location with respect to the trigger pulse. We model the observed
distribution with a convolution product of an exponential decay and
a Gaussian distribution,
\begin{equation}\label{eq:modelfunction}
d_i(t)={1\over2\tau_e} e^{-{\tau_G^2\over4\tau_e^2}}\cdot
e^{t-t_0\over\tau_e}{\rm erfc}\left({t-t_0\over\tau_G}\right)
\end{equation}
The fit values for the temporal offset $t_0$ and the exponential and
Gaussian decay constants $\tau_e, \tau_G$ for the four detectors
$i=1,2,3,4$ are summarized in table~\ref{tab:fitresults}. While the
difference between $\tau_e$ and $\tau_G$ differ maximally by 38\,ps
and 20\,ps, respectively, the time offsets $t_0$ can differ up to
240\,ps between detectors 2 and 4. The physical origin of this
difference could be attributed to differences in the electrical
delays for the different detectors on the order of a few cm on the
circuit board layouts, and to different absolute pulse heights of
the detected breakdown currents due to different parasitic
capacities for the different diodes.
\begin{table}
\caption{Extracted model parameters for the time distributions of
the different photodetectors with their statistical uncertainties.}
\begin{center}
\begin{tabular}{c||c|c|c}
Detector $i$&$t_0$\,(ps)&$\tau_e$\,(ps)&$\tau_G$\,(ps)\\
\hline\hline 1&$1138\pm7$&$395\pm7$&$288\pm4$\\ \hline
2&$1356\pm6$&$433\pm7$&$279\pm4$\\ \hline
3&$1248\pm4$&$409\pm5$&$292\pm3$\\ \hline
4&$1117\pm7$&$415\pm7$&$302\pm4$
\end{tabular}
\label{tab:fitresults}
\end{center}
\end{table}

\section{Information extraction}
An eavesdropper can exploit these differences in the detector
responses $d_i$, and obtain information on the secret key by
listening in the publicly communicated detection times. The
knowledge in principle attainable by the eavesdropper is quantified
by the mutual information $I(X;T)$ between the time distribution of
detector clicks (publicly revealed) and the bits composing the
secret key:
\begin{equation}
I(X;T)=H(X)+H(T)-H(X,T) \label{eq:mutinfo}
\end{equation}
There, $X$ represents the distribution of logical 0 and 1, and $T$
is the distribution of detection times. The entropies and joint
entropies of the distributions are given by
\begin{eqnarray*}
H(T)&=&-\int \bar{d}(t)\log_2[\bar{d}(t)]\,{\rm d}t
\\
H(X)&=&-\sum_x p^0(x)\log_2[p^0(x)]\\
H(X,T)&=&-\sum_x\int p(x,t)\log_2[p(x,t)]\,{\rm d}t\\
&=&-\sum_x\int p^0(x)d_x(t)\log_2[p^0(x)d_x(t)]\,{\rm d}t
\end{eqnarray*}
where $\bar{d}(t)=\sum_xp^0(x)d_x(t)$ is the probability of a click
occurring at time $t$ for the ensemble of detectors, and $d_x(t)$
the probabilities of a click at a particular time $t$ for a detector
corresponding to logical value $x\in \{0,1\}$. In most protocols,
the prior distribution of logical values is balanced such that
$p^0(0)=p^0(1)=0.5$.

If we bin the detector results in the manner most favorable to the
eavesdropper by assigning detectors (1,2) to one basis, (3,4) to the
other basis, and taking detectors groups (1, 3) and (2, 4) to
represent 0 and 1, the average extractable information is
$3.8\pm0.38\%$.
\begin{figure}
\begin{center}
\includegraphics{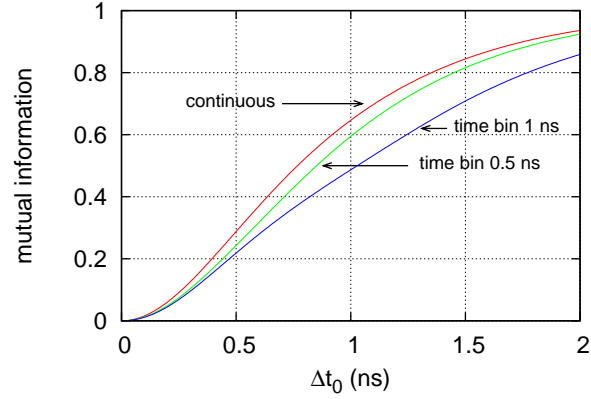}
\end{center}
\caption{Eavesdropper's information on the secret bit as a function
of delay
  $\Delta t_0$ between detector timing distributions with identical
  shapes. The three curves represent different levels of
  discretization of the data. The top curve corresponds to the
  continuous distribution and the subsequent are for 0.5 ns and
  1 ns time bins. As expected, with an increasing time bin there is less information available for the eavesdropper. For
$\Delta t_0$ as small as 0.5\,ns the eavesdropper will gain
  access to more than a quarter of the ``secret key''.}
\label{fig:timeshift}
\end{figure}
It is worth considering in detail how the distinguishability of the
distributions comes about, and how quickly the eavesdropper
knowledge of the key changes. Figure~\ref{fig:timeshift} shows the
eavesdropper's knowledge of the secret bit for two distributions
$d_0(t),d_1(t)$ with the same $\tau_e=400$\,ps, $\tau_G=290\,$ps,
but with different relative delays $\Delta t_0$.  Detectors that are
uncompensated by as little as $\Delta t_0=500\,{\rm ps}$  will give
the eavesdropper access to more than 25\% of the ``secret'' key.
Since a small relative delay is not visible in the usual
experimental setups which employ coincidence windows between 1 and
20\,ns~\cite{poppe:04,resch:05,peng:05,marcikic:06}, it requires an
additional effort to make sure that this leakage channel is closed.

The solution to this particular side channel is not complex, the
timing information should be characterized and the delays equalized,
randomized or the precision truncated such that the potential
information leakage is below a certain threshold. Quantum
cryptography protocols can then deal with this in the same way they
deal with errors, by applying an appropriate amount of privacy
amplification~\cite{bennett:88}. In every real experiment the timing
information is communicated with a finite precision that could be
adjusted for this purpose. Figure 4 shows the effect of discretizing
the time information into 0.5 ns and 1 ns time bins (a typical
experimental value of $\approx150$~ps gives a negligible difference
with the continuous distribution). As expected, the eavesdropper's
information is reduced as the bin width increases. Somewhat
counterintuitively there is still a strong leakage even at bin sizes
comparable to the width of the distribution $d(t)$; furthermore
there is a penalty in the form of increased background. For our
particular device, the main distinguishing feature is the time
offset. If this is compensated for (i.e. made identical for all
detectors), and applying the same procedure as before to obtain the
leakage to an eavesdropper given the probability distributions, we
find the leakage to be around 0.3\%.

It is reasonable to ask whether this problem affects ``prepare and
measure'' protocols as well. A typical BB84 QKD system based on weak
coherent pulses has a synchronous operation, and the detector side
will locally determine whether the detected event falls in the right
part of the timing frame to be counted as genuine. This binary
decision will not provide information to the eavesdropper from the
detector side. However, the problem has just been displaced from the
detectors to the emitters: if the states to be sent are prepared by
different physical devices, their temporal response needs to be
charaterized, and the possible information leakage should be
evaluated with a similar analysis.

\section{Conclusions}
Quantum cryptography is slowly leaving the purely academic
environment and starting to appear in commercial
products~\cite{magiqidq}. The theoretical aspects of its security
are a very active research area but comparatively little has been
done in terms of scrutinizing the practical systems. However, there
is increasing interest in looking at the side channels arising from
the physical realization in practical systems (see recent work by
Zhao et al.~\cite{zhao:07} for an attack on a commercial product
based on a proposal by Makarov et al.~\cite{makarov:06}). We have
shown here how some of the information publicly revealed by the
communicating parties in reasonable mature implementations, may lead
to a large proportion of the key becoming insecure.

\section*{Acknowledgements}
The authors thank Artur Ekert for useful discussions. This work was
partly funded by DSTA in Singapore.
\end{document}